\documentclass[showpacs,preprintnumbers,twocolumn]{revtex4-1}
\usepackage{graphicx,amsmath,amssymb,dcolumn}

\begin{document}

\title{Ensemble and Trajectory Thermodynamics: A Brief Introduction.}

\author{Christian Van den Broeck} 
\affiliation{Hasselt University - B-3590 Diepenbeek, Belgium}
\author{Massimiliano Esposito}
\affiliation{Complex Systems and Statistical Mechanics, University of Luxembourg, L-1511 Luxembourg, Luxembourg}


\begin{abstract}
We revisit stochastic thermodynamics for a system with discrete energy states in contact with a heat and particle reservoir. 
\end{abstract}

\maketitle

\section{Introduction}

Over the last few years, it has become clear that one can extend thermodynamics, which is traditionally confined to the description of equilibrium states of macroscopic systems or to the transition between such states, to cover the  nonequilibrium dynamics of  small scale systems. This extension has been carried out at several levels of description including systems described by discrete and continuous Markovian and non-Markovian stochastic dynamics, by classical Hamiltonian and quantum Hamiltonian dynamics and by thermostatted systems. These developments can be seen, on one side, as extending the work of Onsager and Prigogine \cite{OnsagerReci1, OnsagerReci2, PrigoThermo, GrootMazur} by including microscopic dynamical properties into the far from equilibrium irreversible realm. On the other side, they have led to the reassessment of the cornerstone of thermodynamics, namely the second law of thermodynamics, which is replaced by a much deeper symmetry relation, embodied in the integral and detailed fluctuation theorems. On the more practical side, the new formulation allows to address new questions, either related to nonequilibrium properties, such as efficiency at maximum power or information to work conversion, or relating to the thermodynamic description of small systems, e.g., the discussion of Brownian motors and refrigerators or the efficiency of quantum dots and other small scale devices. In this paper, we present a brief introduction to stochastic thermodynamics. We refer to the literature for more advanced reviews \cite{EvansMorrisB, BustamanteRitortPD, EspositoReview, HanggiFTRMP11, JarzynskiRev11, Seifert12Rev, VdBRev12, QianPR12a, QianPR12b}.

\section{Ensemble thermodynamics }

We consider a system, with discrete non-degenerate states, in contact with a single (ideal, non-dissipative) heat and particle reservoir at temperature $T$ ($\beta=1/(k_B T)$) and chemical potential $\mu$. The states are  identified by an index  $m$, with corresponding energy $\epsilon_m$ and particle number $n_m$. For simplicity we consider a single type of particle. We assume that the system can also exchange work with an (ideal, non-dissipative) work source which controls its energy levels $\epsilon(\lambda)$ via an time-dependent control parameter $\lambda=\lambda(t)$. The particle number in a given state is however supposed to be fixed.

In the ensemble picture, the state of the system is described by a probability distribution $P_m$ to be in the state $m$, with $\sum_m P_m=1$. Note that this distribution does not have to be of the equilibrium form, so that ``traditional equilibrium concepts'', such as temperature and chemical potential need not exist for the system. They are however well defined for the ideal reservoir, and when appearing in the formulas below, $T$ and $\mu$ refer to this reservoir. The time evolution of the state is described by a Markovian master equation:
\begin{eqnarray}
d_t P_m=\sum_{m'} W_{m,m'}P_{m'}.
\end{eqnarray}
Here $W_{m,m'}$ is the  probability per unit time  to make a transition from state $m'$ to $m$. We use the shorthand notation with diagonal elements defined as $W_{m,m}=-\sum_{m'\neq m} W_{m',m}$. Alternatively:
\begin{eqnarray}
\sum_{m} W_{m,m'}=0,
\end{eqnarray}
a property that guarantees the conservation of normalization. 
The transition rates have to satisfy an additional property. In the steady state, the system is at equilibrium with the reservoir. Statistical physics prescribes that the steady state distribution is given by the grand canonical equilibrium distribution $P^{eq}_m$ \cite{ReichlBook}:
\begin{equation}\label{eqd}
P^{eq}_m=\exp\{-\beta(\epsilon_m-\mu n_m -\Omega^{eq})\}.
\end{equation}
The (equilibrium) grand potential $\Omega^{eq}$ follows from the normalization of $P^{eq}$:
\begin{equation}
\exp\{-\beta\Omega^{eq}\}=\sum_m \exp\{-\beta(\epsilon_m-\mu n_m )\}.
\end{equation}
The crucial property that is required from the rates is the so-called condition of detailed balance, i.e., at equilibrium every  transition, say from $m$ to $m'$, and its inverse, from $m'$ to $m$, have to be equally likely: 
\begin{eqnarray}
W_{m,m'}{P}^{eq}_{m'}=W_{m',m}{P}^{eq}_{m}.
\end{eqnarray}
Combined with the explicit expression of the equilibrium distribution, this gives: 
\begin{eqnarray}\label{W}
k_B\ln{\frac{W_{m',m}}{W_{m,m'}}}=\frac{\epsilon_{m}-\epsilon_{m'}-\mu (n_{m}-n_{m'})}{T}=\frac{q_{m,m'}}{T}.
\end{eqnarray}
${q_{m,m'}}$ is the ``elementary'' heat absorbed by the system to make the transition from $m'$ to $m$. We stress that in the presence of driving, which is shifting the energy levels in time, this relation is supposed to hold at each moment in time, hence the rates also become time-dependent. This condition will be crucial to obtain the correct formulation of the second law.

We next introduce the basic state functions -quantities  that depend on probability distribution $P_m$ of the system, but not on the way this distribution was achieved- namely the ensemble-averaged and in general nonequilibrium values of energy, particle number and entropy:
\begin{eqnarray}\label{E}
E&=&\sum_m \epsilon_m P_m= \langle \epsilon_m \rangle ,\\
N&=&\sum_m n_m P_m=\langle n_m \rangle  \label{N} ,\\
S&=&-k_B \sum_m P_m \ln P_m=\langle -k_B \ln P_m \rangle.\label{S}
\end{eqnarray}

It is clear from the above formulas that these state variables can change due to two different mechanisms: a change in occupation of the levels, i.e., a modification  of $P_m$, or a   shift of the energy levels, i.e., a change of the energy level $\epsilon_m$. In particular, the rate of energy change is given by:
\begin{eqnarray}\label{fl}
d_t E&=&\sum_m \{\epsilon_m d_t P_m+d_t \epsilon_m P_m\}\\
&=& \dot{Q}+\dot{W}_{chem}+\dot{W}
\end{eqnarray}
This decomposition reproduces the first law. Work rate (power) $\dot{W}$, particle flow $d_t N$ and heat flow $\dot{Q}$ are given by:
\begin{eqnarray}
\dot{W} &=& \sum_m d_t \epsilon_m P_m = d_t \lambda \; d_{\lambda} E ,\\
d_t N&=&\sum_m n_m d_t P_m \label{Nr} ,\\
\dot{Q} &=& \sum_m \epsilon_m d_t P_m- \dot{W}_{chem} \label{Qr}.
\end{eqnarray}
with the chemical work rate $\dot{W}_{chem}$
\begin{eqnarray}
\dot{W}_{chem} &=&  \mu d_t N \label{Wcr}.
\end{eqnarray}
In words, work is the result of the energy shift of an occupied state. Heat and chemical work correspond to transitions between states, implying a change in occupation probability. As is well known, heat and work are not state functions (or the difference of state functions) as they depend on the way the transition between states is carried out. We also mention for later use the following expression for the heat flux, cf. (\ref{Nr},\ref{Qr},\ref{Wcr}):
\begin{equation}\label{heatflux}
\dot{Q}= \sum_m (\epsilon_m-\mu n_m) d_t P_m.
\end{equation}

Turning to the  second law, we will show that the above definition of nonequilibrium entropy is a proper choice that reduces to the standard thermodynamic entropy at equilibrium, but with the additional advantage that it preserves -in nonequilibrium- the basic features of the second law, namely the relation between heat and entropy exchange and the positivity of the entropy production. Explicitly:
\begin{eqnarray}
d_t S&=&\dot{S}_e+\dot{S}_i\label{ES} ,\\
\dot{S}_e&=&\dot{Q}/T \label{EE} ,\\
 \dot{S}_i&\geq& 0 \label{EP}.
\end{eqnarray}
 The proof goes as follows:
\begin{eqnarray}\label{ssl}
d_t S&=&-k_B\sum_{m}d_t P_{m}\ln{P}_{m}=-k_B \sum_{m,m'}W_{m,m'}{P}_{m'}\ln{P}_{m}\nonumber\\
&=&-k_B \sum_{m,m'}W_{m,m'}{P}_{m'}\ln\frac{{P}_{m}}{{P}_{m'}}\nonumber\\
&=&\frac{k_B}{2}\sum_{m,m'}\left(W_{m,m'}{P}_{m'}-W_{m',m}{P}_{m}\right)\ln\frac{{P}_{m'}}{{P}_{m}}\nonumber\\
&=&\frac{k_B}{2}\sum_{m,m'}\left(W_{m,m'}{P}_{m'}-W_{m',m}{P}_{m}\right)\ln\frac{W_{m,m'}{P}_{m'}}{W_{m',m}{P}_{m}}\nonumber\\
&+& \frac{k_B}{2}\sum_{m,m'}\left(W_{m,m'}{P}_{m'}-W_{m',m}{P}_{m}\right)\ln\frac{W_{m',m}}{W_{m,m'}}.
\end{eqnarray}
We thus identify the entropy production and entropy flow as:
\begin{eqnarray}\label{ss0}
&&\hspace{-0.45cm} \dot{S}_i=\frac{k_B}{2}\sum_{m,m'}\left(W_{m,m'}{P}_{m'}-W_{m',m}{P}_{m}\right)\ln\frac{W_{m,m'}{P}_{m'}}{W_{m',m}{P}_{m}},\\
&&\hspace{-0.45cm} \dot{S}_e= \frac{k_B}{2}\sum_{m,m'}\left(W_{m,m'}{P}_{m'}-W_{m',m}{P}_{m}\right)\ln\frac{W_{m',m}}{W_{m,m'}},
\end{eqnarray}
or:
\begin{eqnarray}\label{ssl2}
\dot{S}_i&=&\sum_{m>m'}J_{m,m'}X_{m,m'},\\
\dot{S}_e&=&\sum_{m>m'}J_{m,m'}\frac{q_{m,m'}}{T},
\end{eqnarray}
with
\begin{eqnarray}\label{JX}
J_{m,m'}&=&W_{m,m'}{P}_{m'}-W_{m',m}{P}_{m},\\
X_{m,m'}&=&k_B\ln\frac{W_{m,m'}{P}_{m'}}{W_{m',m}{P}_{m}},\\
q_{m,m'}&=&\epsilon_{m}-\epsilon_{m'}-\mu (n_{m}-n_{m'}).
\end{eqnarray}
$J_{m,m'}$ is the net rate of transitions from $m'$ to $m$ and $X_{m,m'}$ the corresponding thermodynamic force. $\dot{S}_i$ is clearly non-negative. In fact, a stronger property holds: it is  the sum of contributions due to all pairwise transitions between different states $m$ and $m'$, and each of these terms in non-negative. Concerning $\dot{S}_e$, one  finds using (\ref{W},\ref{heatflux}):
\begin{eqnarray}
\dot{S}_e &=& {k_B}\sum_{m,m'}W_{m,m'}{P}_{m'}\ln\frac{W_{m',m}}{W_{m,m'}} \\
&=& \sum_{m,m'}W_{m,m'}{P}_{m'}\frac{\epsilon_{m}-\epsilon_{m'}-\mu (n_{m}-n_{m'})}{T}\\
&=& \sum_{m,m'}W_{m,m'}{P}_{m'}\frac{\epsilon_{m}-\mu n_{m}}{T}\\
&=& \sum_{m}d_t P_{m}\frac{\epsilon_{m}-\mu n_{m}}{T}=\frac{\dot{Q}}{T},
\end{eqnarray}
which is indeed the expected expression for the entropy flow.
 
Besides energy, particle number and entropy, it is  convenient to introduce another state function, namely the grand potential:
\begin{eqnarray}\label{gp}
\Omega=E-T S-\mu N.
\end{eqnarray}
One immediately finds that:
\begin{eqnarray}
d_t \Omega=d_t E-T d_t S-\mu d_t N=\dot{W}-T\dot{S}_i
\end{eqnarray}
or
\begin{eqnarray}\label{gpep}
T\dot{S}_i=\dot{W}-d_t \Omega \geq 0.
\end{eqnarray}
As a consequence the rate of decrease $-d_t\Omega$ in grand potential is an upper bound for the corresponding output power $-\dot{W}$. 

The above description provides a generalization of the usual equilibrium thermodynamics to far from equilibrium states. It is reassuring that at equilibrium it reproduce the properties of the equilibrium state. By filling in the equilibrium expression for the probability distribution (\ref{eqd}) in (\ref{E},\ref{N},\ref{S},\ref{gp}), one finds the corresponding results for the energy, number of particles, entropy and grand potential, $E^{eq}$, $N^{eq}$, $S^{eq}$, $\Omega^{eq}$. At equilibrium, (\ref{gp}) thus reproduces the equilibrium Euler relation $\Omega^{eq}=E^{eq}-T S^{eq}-\mu N^{eq}$. Note furthermore that the knowledge of the equilibrium grand potential as a function of $T$, $\lambda$ and $\mu$, $\Omega^{eq}=\Omega^{eq}(T,\lambda,\mu)$ provides a fundamental relation, i.e., a full characterization of the system. The energy-particle spectrum  that characterizes the system can be recovered through an inverse Laplace transform \cite{ReichlBook}. This is obviously no longer true for the nonequilibrium potential $\Omega$. We however point out the following revealing property: the nonequilibrium potential is always larger than its equilibrium value:
\begin{eqnarray}\label{gpi}
\Omega-\Omega^{eq} &=& k_B T I ,\\
I=D(P||P^{eq}) &=& \sum_m P_m \ln \frac{P_m}{P^{eq}_m} \geq 0.
\end{eqnarray}
Here $D(P||P^{eq})$ is the relative entropy or Kullback Leibler distance \cite{CoverThomas} between the distributions $P$ and $P^{eq}$. Combined with the second law under the form (\ref{gpep}), we conclude that any nonequilibrium state has the potential to generate an output work of at most $k_B T I$, upper limit reached for a reversible scenario (zero entropy production).

We finally consider a quasi-static transformation generated by the work source. In this case the equilibrium shape of the distribution is preserved in time, $P_m=P_m^{eq}(t)$. Hence, as the energy levels are shifted via work, a corresponding instantaneous redistribution over the levels has to take place, with a concomitant heat and particle exchange.  One immediately finds that for such a transformation:
\begin{eqnarray}
d_t S&=&-k_B \sum_m d_t P^{eq}_m \ln P_m^{eq}\\
&=&\sum_m d_t P^{eq}_m \frac{\epsilon_m-\mu n_m-\Omega^{eq}}{T} \\
&=&\frac{\dot{Q}}{T}=\dot{S}_e.
\end{eqnarray}
This also implies $\dot{S}_i=0$ and $\dot{W}=d_t\Omega^{eq}$. These relations reproduce the familiar thermodynamic statement for quasi-static processes: $d_t S=\dot{Q}/{T}$.
   
\section{Multiple reservoirs}

The case of multiple energy and particle reservoirs is of obvious theoretical and technological interest. On the theory side, we mention the study of nonequilibrium steady states. With respect to applications, engines, pumps and refrigerators all involve contact with multiple reservoirs. 
We briefly explain how the above formalism can be extended to cover this situation.
The main new ingredient is the fact that the transitions in the system are now due to the coupling with the different reservoirs, which we identify by the index $\nu$ (e.g., temperature $T^\nu$, chemical potential $\mu^\nu$, etc.). We assume that these contacts do not interfere, hence:
\begin{eqnarray}
W_{m,m'}=\sum_\nu W^\nu_{m,m'},
\end{eqnarray}
with $W^\nu$ the transition matrix due to coupling with reservoir $\nu$. This latter satisfies detailed balance
\begin{eqnarray}
W^\nu_{m,m'}{P}^{eq,\nu}_{m'}=W^\nu_{m',m}{P}^{eq,\nu}_{m}
\end{eqnarray}
with respect to the equilibrium distribution ${P}^{eq,\nu}$ imposed by reservoir $\nu$ at each moment in time (i.e. for the prevailing energy spectrum $\epsilon_m=\epsilon_m(t)$):
\begin{eqnarray}\label{eqd2}
P^{eq,\nu}_m=\exp\{-\beta(\epsilon_m-\mu^\nu n_m -\Omega^{eq,\nu})\},\\
\exp\{-\beta\Omega^{eq,\nu}\}=\sum_m \exp\{-\beta^\nu(\epsilon_m-\mu^\nu n_m )\}.
\end{eqnarray}
We can now write:
\begin{eqnarray}
d_t P_m&=&\sum_\nu \dot{P}_m^\nu ,\\ 
\dot{P}_m^\nu&=&\sum_{m'} W^\nu_{m,m'}P_{m'}.
\end{eqnarray}
Hence, in all of the above the formulas where the derivative of the probability appears, one can identify the separate contributions of the different reservoirs. In particular, the first law (\ref{fl}) remains of course valid, but particle, heat and chemical work flux can be separated into contributions from each reservoir:
\begin{eqnarray}
d_t N &=& \sum_\nu \dot{N}^\nu,\\
\dot{Q} &=& \sum_\nu \dot{Q}^\nu,\\
\dot{W}_{chem} &=& \sum_\nu \dot{W}_{chem}^\nu,
\end{eqnarray}
with
\begin{eqnarray}
\dot{N}^\nu&=&\sum_{m} n_m \dot{P}_m^\nu \label{mNr} ,\\
\dot{Q}^\nu &=& \sum_{m} (\epsilon_m-\mu^\nu n_m) \dot{P}_m^\nu ,\\
\dot{W}_{chem}^\nu &=& \sum_{m} \mu^\nu n_m \dot{P}_m^\nu\label{mWcr}.
\end{eqnarray}
For the second law, the derivation is modified as follows:
\begin{eqnarray}
d_t S &&= -k_B\sum_{m}d_t P_{m}\ln{P}_{m}=-k_B\sum_{\nu} \sum_{m,m'}W_{m,m'}^\nu{P}_{m'}\ln{P}_{m}\nonumber\\
&&\hspace{-0.4cm}=\sum_{\nu}\frac{k_B}{2}\sum_{m,m'}\left(W_{m,m'}^\nu{P}_{m'}-W_{m',m}^\nu{P}_{m}\right)\ln\frac{{P}_{m'}}{{P}_{m}}\nonumber\\
&&\hspace{-0.4cm}=\frac{k_B}{2}\sum_{\nu}\sum_{m,m'}\left(W_{m,m'}^\nu{P}_{m'}-W_{m',m}^\nu{P}_{m}\right)\ln\frac{W_{m,m'}^\nu{P}_{m'}}{W_{m',m}^\nu{P}_{m}}\nonumber\\
&&\hspace{-0.2cm}+ \frac{k_B}{2}\sum_{\nu}\sum_{m,m'}\left(W_{m,m'}^\nu{P}_{m'}-W_{m',m}^\nu{P}_{m}\right)\ln\frac{W_{m',m}^\nu}{W_{m,m'}^\nu}.
\end{eqnarray}
One thus finds:
\begin{eqnarray}
\dot{S}_i&=& \sum_{\nu}\dot{S}_i^\nu,\\
\dot{S}_e&=&\sum_{\nu}\dot{S}_e^\nu,
\end{eqnarray}
with
\begin{eqnarray}
\dot{S}_i^\nu&=&\sum_{m>m'}J^\nu_{m,m'}X^\nu_{m,m'},\\
\dot{S}_e^\nu&=&\sum_{m>m'}J^\nu_{m,m'}\frac{q^\nu_{m,m'}}{T^\nu}=\frac{\dot{Q}^\nu}{T^\nu},
\end{eqnarray}
and 
\begin{eqnarray}\label{JX2}
J^\nu_{m,m'}&=&W^\nu_{m,m'}{P}_{m'}-W^\nu_{m',m}{P}_{m},\\
X^\nu_{m,m'}&=&{k_B}\ln\frac{W^\nu_{m,m'}{P}_{m'}}{W^\nu_{m',m}{P}_{m}} ,\\
q^\nu_{m,m'}&=&\epsilon_{m}-\epsilon_{m'}-\mu^\nu (n_{m}-n_{m'}).
\end{eqnarray}

The above formulas allow to investigate the far from equilibrium thermodynamics of small scale systems, for example of quantum dots in contact with leads \cite{EspositoHarbola07, EspoLindVdB_EPL09_Dot, EspKawLindVdBEPL10, EspoStrassSchaBrandPRE13}. Another question that has received a lot of attention is the universal features of the efficiency of such machines when operating at maximum power \cite{VandenBroeckPRL05, SeifertSchmiedlPRL07, Tu08, EspositoPRL09, EspoRuttCleuPRB, EspoKawLindVdB_PRE_10, SeifertPRL11, VdBCleuren12PRL, SeifertBrandnerPRL13, Tu2013}. We finally mention the simplification in the absence of particle transport, achieved by setting in the above formulas all the chemical potentials equal to zero, $\mu^\nu=0, \forall \nu$. 

\section{Trajectory thermodynamics}

In the limit of a very large system (with no long range correlations), one expects by the law of large numbers that the properties are self-averaging, so that an ensemble description in terms of average quantities is sufficient. In  a small scale system, this is no longer the case and the quantities that are measured will vary from one experiment to another. One could of course still apply the above presented ensemble thermodynamics to describe the average outcome upon repeating the experiment many times. Nevertheless, one can  wonder about  the stochastic properties that are revealed at the trajectory level. Furthermore, due to the spectacular progress in nano-technology,  the experimental measurement of such properties are now within reach. But there is an even more important motivation: it turns out, as we  proceed to show, that the trajectory properties  reveal a much deeper formulation of the second law.  We will limit our presentation, for clarity and simplicity of notation, to the case of a single particle and energy reservoir.

We thus consider a small system and focus on its trajectory in the course of time in a single  realization of the experiment. The state of the system at time $t$ is its actual state  $m=m(t)$. The analogues of the above introduced ensemble averaged state functions $E$, $N$, $S$ and $\Omega$ for this particular state are  the stochastic  energy $e$, the stochastic number of particles $n$, the stochastic entropy $s$ introduced by Seifert \cite{Seifert05}, and the stochastic grand canonical potential $\omega$ of this state: 
\begin{eqnarray}\label{e}
e&=& \epsilon_{m(t)}(t), \\
n&=& n_{m(t)}\label{n},\\
s&=&-k_B\ln P_{m(t)}(t),\label{s}\\
\omega&=&e-Ts-\mu n.\label{omega}
\end{eqnarray}
We have assumed that the particle number of a given state is fixed, so that $n_{m(t)}$ has no explicit time dependence. The above definitions are consistent with the ensemble description, $E=\langle e \rangle$, $N=\langle n \rangle$, $S=\langle s \rangle$ and $\Omega=\langle \omega \rangle$, where the averaging bracket refer to an average with respect to the probability distribution $P_{m(t)}(t)$. We will henceforth use the lower case notation to distinguish the stochastic variables from the corresponding ensemble averaged quantities (it should be clear from the context not to confuse the energy $e$ with Euler's number $e=2,718...$).
Note that at the trajectory level, there appears an  essential difference between the  variables $e$ and $n$ on the one hand, and $s$ and $\omega$ on the other hand: the latter retain an ensemble character, trademark of their thermodynamic content, as they depend, not only on the actual state $m(t)$, but also on the probability distribution $P_{m(t)}$. Hence, even though we are monitoring single trajectories, their thermodynamic properties are defined with respect to  the stochastic dynamics of the system under consideration.

In the following it will be convenient, in order to take time-derivatives, to extract the time-dependence on the actual state $m(t)$ with a delta-Kronecker function:
\begin{eqnarray}
f_{m(t)}(t)&=&\sum_m \delta^{Kr}_{m,m(t)} f_{m}(t) ,\\
d_t f_{m(t)}(t)&=&\sum_m \dot{\delta}^{Kr}_{m,m(t)} f_{m}(t)+ \delta^{Kr}_{m,m(t)} d_t f_{m}(t) .
\end{eqnarray}
Note that the contribution in $\dot{\delta}^{Kr}_{mm'}$ correspond to a change of level occupation from state $m'$ to $m$ and will give rise to a Dirac delta function contribution centered at the times $t^*$ of the jumps and with amplitude $f_{m}(t^*)-f_{m'}(t^*)$.

The first law at the trajectory level is obtained by differentiating (\ref{e}) with respect to time :
\begin{eqnarray}\label{dote}
d_t e&=& \dot{q}+\dot{w}_{chem}+\dot{w},\\
\dot{w}&=&\sum_m {\delta}^{Kr}_{m,m(t)} d_t \epsilon_m(t) \label{dotw} ,\\
\dot{q}&=&\sum_m \epsilon_m(t) \dot{\delta}^{Kr}_{m,m(t)} -\dot{w}_{chem} \label{dotq} ,\\
\dot{w}_{chem}&=&\sum_m \mu n_m \dot{\delta}^{Kr}_{m,m(t)}. \label{dotchem}
\end{eqnarray}
The interpretation is essentially the same as in the ensemble description: work corresponds to the shift of an occupied level while heat plus chemical work are the result of a transition between levels.

The second law at the trajectory level is obtained by calculating as follows the time-derivative of the stochastic entropy (\ref{s}):
\begin{eqnarray}\label{dots1}
d_t s&=& -k_B\sum_m \{\dot{\delta}^{Kr}_{m,m(t)} \ln P_{m}(t)+{\delta}^{Kr}_{m,m(t)} \frac {d_t P_{m}(t)}{{P}_{m}(t)}\}\nonumber\\
&=& -k_B\sum_m \{\dot{\delta}^{Kr}_{m,m(t)} \ln \frac{P_{m}(t)}{P^{eq}_m(t)}+{\delta}^{Kr}_{m,m(t)} \frac {d_t P_{m}(t)}{{P}_{m}(t)}\nonumber\\
&&+\dot{\delta}^{Kr}_{m,m(t)} \ln {P^{eq}_m(t)}\}.
\end{eqnarray}
Here we introduced the instantaneous equilibrium distribution $P^{eq}_{m}(t)$, see also (\ref{eqd}):
\begin{equation}\label{eqdt}
P^{eq}_m(t)=\exp\{-\beta[\epsilon_m(t)-\mu n_m -\Omega^{eq}(t)]\}.
\end{equation}
The crucial point is to identify the last term in the r.h.s. of (\ref{dots1}) as the stochastic entropy flow, using (\ref{dotq},\ref{dotchem}):
\begin{eqnarray}\nonumber
\dot{s}_e&=&-k_B\sum_m \dot{\delta}^{Kr}_{m,m(t)} \ln {P^{eq}_{m}(t)}\\
&=&\frac{1}{T} \sum_m \dot{\delta}^{Kr}_{m,m(t)}  \{{\epsilon_{m}(t)}-\mu n_{m}\}\nonumber\\
&=&\frac{\dot{q}}{T}\label{dotse}.
\end{eqnarray}
From the decomposition 
\begin{eqnarray}\label{dots2}
d_t s= \dot{s}_i+\dot{s}_e,
\end{eqnarray}
first obtained by Seifert in Ref. \cite{Seifert05}, we conclude that the stochastic entropy production is given by
\begin{eqnarray}
\hspace{-0.35cm} \dot{s}_i = -k_B\sum_m \{\dot{\delta}^{Kr}_{m,m(t)} \ln \frac{P_{m}(t)}{P^{eq}_m(t)}+{\delta}^{Kr}_{m,m(t)} \frac {d_t P_{m}(t)}{{P}_{m}(t)}\} \label{dots}.
\end{eqnarray}
It consists of two different terms. The second one in the r.h.s. corresponds to  a smooth contribution: it is positive or negative depending on whether the actual state becomes less or more probable in time. The first term in the r.h.s. gives discontinuous contributions, appearing at the instances when the actual state of the system changes. Note that neither of the above contributions has a definite sign. 

By averaging the result (\ref{dotse}) for entropy flow with $P_{m(t)}(t)$, one recovers its ensemble version (\ref{EE}). To do the same for the entropy production requires a little bit more work. By summation over all the actual states $m(t)=m'$ (i.e. this index becomes a dummy summation variable)  one finds that the second term in the stochastic entropy production (\ref{dots}) averages out to zero. As for the first term,  we note that the stochastic entropy production jumps by an amount $k_B \ln \frac{P_{m'} P^{eq}_{m}}{P^{eq}_{m'} P_{m}}$ for a change in state from $m'$ to $m$. The probability per unit time for such a transition is $W_{m,m'}P_{m'}$. The average $\langle \dot{s}_i \rangle$ of the stochastic entropy production thus becomes (note that the averaging brackets now refer to an average with respect to $W_{m,m'}P_{m'}$, and that we have dropped for simplicity of notation the explicit $t$ dependence):
\begin{eqnarray}\label{dotsav}
\dot{S}_i &=& k_B\sum_{m,m'} W_{m,m'}P_{m'} \ln \frac{P_{m'} P^{eq}_{m}}{P^{eq}_{m'} P_{m}}\\ 
&=& k_B\sum_{m,m'} W_{m,m'}P_{m'} \ln \frac{W_{m,m'}P_{m'}}{W_{m',m}  P_{m}} \nonumber\\
&&\hspace{-0.55cm} =\frac{k_B}{2}\sum_{m,m'} \{W_{m,m'}P_{m'}-W_{m',m}P_{m}\} \ln \frac{W_{m,m'}P_{m'}}{W_{m',m}  P_{m}} \geq 0. \nonumber
\end{eqnarray}
This is indeed the  expression for the ensemble entropy production given earlier. 

 
Turning to the stochastic grand potential (\ref{omega}), one finds by a combination of the stochastic first and second law, very much in the same way as for the ensemble average, that (compare with (\ref{gpep})):
\begin{eqnarray}\label{epgp}
T\dot{s}_i=\dot{w}-d_t \omega.
\end{eqnarray}
We next make the important observation that, when the probability distribution has the equilibrium form $P_m=P_m^{eq}$, and in particular for a quasi-static process $P_m=P_m^{eq}(t)$, cf. (\ref{eqdt}), the stochastic grand potential given in (\ref{omega}) becomes independent of the state, and reduces to the ensemble average instantaneous equilibrium expression (using the explicit expression for the stochastic entropy given in (\ref{s})):
\begin{eqnarray}\label{gpe}
\omega^{eq}(t)=\Omega^{eq}(t).
\end{eqnarray}
For a transition between an initial and a final equilibrium state (but with no conditions on the state of the system in the intermediate process), one can thus write by integration of (\ref{epgp}):
\begin{eqnarray}\label{entropywork}
T\Delta_i{s}={w}-\Delta{\Omega^{eq}},
\end{eqnarray}
with the important consequence that the statistical properties of entropy production $\Delta_i{s}$ and work $w$ are the same, apart from a shift by $\Delta{\Omega^{eq}}$ and rescaling by $T$.
Finally, the difference of equilibrium and  nonequilibrium grand potential can be written as follows:
\begin{eqnarray}\label{gpi2}
\omega-\omega^{eq}&=&k_B T i ,\\
i&=&  \ln\frac{P_m}{P^{eq}_m},
\end{eqnarray}
which is the stochastic analogue of (\ref{gpi}).

\section{Path description}\label{TrajLevel}

The expression (\ref{dots}) for the entropy production is not physically nor mathematically transparent. A  more revealing expression is obtained by considering the cumulative entropy production $\Delta_i{s}$ along a trajectory. We represent a system trajectory by $\bold{m}$, refering to the state of the system, $m(t)$, as a function of time $t=[t_i=0,t_f=\tau]$. The corresponding probability for such a trajectory will be denoted by $\bold{P}(\bold{m})$.  We also need to define an ``inverse experiment'', corresponding to reversing the time-dependence of the ``driving'', i.e., of the transition rates, {\it while using as starting probability for the states in the inverse experiment, the final distribution of the states in the forward experiment}. We will denote by superscript ``tilde'' the quantities related to the reverse experiment, for example $\tilde{t}=\tau-t$, $\tilde{t}_i=\tau-{t}_f=0$, $m(t)=\tilde{m}(\tilde{t})$, $P_{m(t_f)}(t_f)=\tilde{P}_{\tilde{m}(\tilde{t}_i)}(\tilde{t}_i)$, etc.. $\tilde{\bold{m}}$ refers to the time-inverse trajectory of $\bold{m}$ and ${\tilde{\bold{P}}(\tilde{\bold{m}})}$ to the probability for the time-reversed path in the time-reversed experiment. We will now prove the following main result. The cumulated entropy production   $\Delta_i{s}=\Delta_i{s}(\bold{m})$ along a forward trajectory $\bold{m}$ is the logratio of the probabilities to observe this trajectory in forward and backward experiment, respectively:
\begin{eqnarray}\label{ept}
\Delta_i{s}(\bold{m})=k_B \ln\frac{\bold{P}(\bold{m})}{\tilde{\bold{P}}(\tilde{\bold{m}})}.
\end{eqnarray}
Note that the probability for a trajectory is in fact a probability density defined in the function space of trajectories, but as only ratio's of such quantities appear (and the Jacobian for the transformation to the time-inverse variables is one), the above expression is well defined.
The proof is surprisingly simple. One can identify three contributions to the probability for a trajectory. First, we have the starting probabilities for direct and reverse paths, $P_{m(t_i)}(t_i)$ and $\tilde{P}_{m(\tilde{t}_i)}(\tilde{t}_i)=P_{m(t_f)}(t_f)$ (the latter by construction), respectively. The logratio of these quantities gives the stochastic entropy change of the forward trajectory $\Delta s(\bold{m})=k_B\ln P_{m(t_f)}(t_f)-k_B\ln P_{m(t_i)}(t_i)$. Second, we have the probabilities to make jumps, say from $m$ to $m'$ in the forward process and hence from $m'$ to $m$ in the reverse experiment. The logratio of these probabilities is $\ln{\frac{W_{m',m}}{W_{m,m'}}}=\frac{-q_{m',m}}{k_B T}$ (evaluated at the time of the jump), which is the heat $-q_{m',m}$ {\it leaving} the system divided by the $k_B$ times temperature. When summed over all transitions (times $k_B$), it gives $-\Delta_e s(\bold{m})$. Finally the probabilities for not making jumps are at every instant of time the same in the forward and reverse experiments, so these contributions cancel out. We conclude that the logratio of the path probabilities is $\Delta s(\bold{m})-\Delta_e s(\bold{m})=\Delta_i s(\bold{m})$.\\

\section{Fluctuation and work theorem}\label{TrajLevel2}

The cumulated entropy production $\Delta_i{s}$ is a random variable: it has the value specified in  (\ref{ept}) when the system has followed the specified trajectory $\bold{m}$, which happens with probability $\bold{P}(\bold{m})$. The resulting probability for $\Delta_i{s}$ is given by the following path integral:
\begin{eqnarray}\label{probept}
P(\Delta_i{s})=\int_\bold{m}\; d\bold{m}\; \delta\left(\Delta_i{s}-k_B \ln\frac{\bold{P}(\bold{m})}{\tilde{\bold{P}}(\tilde{\bold{m}})}\right)\bold{P}(\bold{m}).
\end{eqnarray}
Due to the very specific structure of $\Delta_i{s}$, namely the fact that it is a  log-ratio of probabilities, one can perform the following trick:   
\begin{eqnarray}\label{probept}
P(\Delta_i{s})&=&\int_\bold{m} \; d\bold{m}\;\delta\left(\Delta_i{s}-k_B \ln\frac{\bold{P}(\bold{m})}{\tilde{\bold{P}}(\tilde{\bold{m}})}\right)\bold{P}(\bold{m}) \nonumber\\
=e^\frac{\Delta_i{s}}{k_B} &&\int_\bold{m} \; d\bold{m}\;\delta\left(\Delta_i{s}-k_B \ln\frac{\bold{P}(\bold{m})}{\tilde{\bold{P}}(\tilde{\bold{m}})}\right){\tilde{\bold{P}}(\tilde{\bold{m}})} \nonumber\\
=e^\frac{\Delta_i{s}}{k_B}&&\int_{\tilde{\bold{m}}}\;d\tilde{\bold{m}}\; \delta\left(-\Delta_i{s}-k_B \ln\frac{\tilde{\bold{P}}(\tilde{\bold{m}})}{\bold{P}(\bold{m})} \right) {\tilde{\bold{P}}(\tilde{\bold{m}})}\nonumber\\
= \exp\frac{\Delta_i{s}}{k_B} &&\tilde{P}(-\Delta_i{s}),
\end{eqnarray}
where we have used the fact that the Jacobian for the transformation from $\bold{m}$ to $\tilde{\bold{m}}$ is equal to one.  This result is usually written under the following form:
\begin{equation}\label{dft}
\frac{P(\Delta_i{s})}{\tilde{P}(-\Delta_i{s})}=\exp\frac{\Delta_i{s}}{k_B},
\end{equation}
also known as the detailed fluctuation theorem \cite{EspositoVdBPRL10}. In words: the probability for a stochastic entropy increase (in the forward dynamics) is exponentially more probable than that of a corresponding decrease (in the reverse dynamics). At this point it is important to clarify the meaning of: 
\begin{equation}
\tilde{P}(-\Delta_i{s})=\int_{\tilde{\bold{m}}}\;d\tilde{\bold{m}}\; \delta\left(-\Delta_i{s}-k_B \ln\frac{\tilde{\bold{P}}(\tilde{\bold{m}})}{\bold{P}(\bold{m})} \right) {\tilde{\bold{P}}(\tilde{\bold{m}})}.
\end{equation}
$\tilde{P}(\tilde{\bold{m}})$ is obviously the probability density to observe in the reverse dynamics a trajectory $\tilde{\bold{m}}$ whose inverse trajectory ${\bold{m}}$ has an entropy production  $\Delta_i{s}$. It is quite natural to wonder what is the relation, if any, with the entropy production $\Delta_i{\tilde{s}}(\tilde{\bold{m}})$ of the reverse trajectory $\tilde{\bold{m}}$ in the reverse scenario. 
By applying (\ref{ept}) to the reverse process one finds that the corresponding entropy production reads:
\begin{eqnarray}\label{sepr}
\Delta_i\tilde{s}(\tilde{\bold{m}})=k_B \ln\frac{\tilde{\bold{P}}(\tilde{\bold{m}})}{ \tilde{\tilde{\bold{P}}}(\tilde{\tilde{\bold{m}}}) }.
\end{eqnarray}
It is  obvious that $\tilde{\tilde{\bold{m}}}=\bold{m}$, and it is tempting to assume that $\tilde{\tilde{\bold{P}}}=\bold{P}$, implying $\Delta_i\tilde{s}(\tilde{\bold{m}})=-\Delta_i s(\bold{m})$. Indeed the transition rates of the doubly tilded stochastic process (twice time-inversion of the driving) are again the original ones $\tilde{\tilde{\bold{W}}}=\bold{W}$,  so that ${\bold{P}}$ and $\tilde{\tilde{\bold{P}}}$ obey the same master equation. But  $\tilde{\tilde{\bold{P}}}=\bold{P}$ also requires that their initial distributions coincide. This is in general not the case. Indeed, we recall that - for (\ref{sepr}) to correspond to the entropy production in the reverse process - the initial probability distribution of  $\tilde{\tilde{\bold{P}}}$ has to be the final probability ${\tilde{\bold{P}}}$. In general the latter distribution  will  be different from the initial distribution of the forward process so that ${\bold{P}}\neq \tilde{\tilde{\bold{P}}}$. Hence, there is in general no simple relation between the entropy production of forward and backward processes, and the fluctuation theorem is a statement only about the forward entropy production. There are however several cases of interest where the initial conditions match. One prominent situation is for the system starting and ending in a stationary state. This is obviously the case for nonequilibrium steady states (same stationary state all the time), but also if both before and after the time-dependent perturbation, the system has enough time to relax to the corresponding steady state. 
A particularly interesting special case of this type arises if we consider a system starting and ending in an equilibrium state. Indeed  the stochastic entropy is then directly related to the work performed on the system, cf. (\ref{entropywork}), and the statistical properties of the entropy production carry over to the work, leading to the so-called Crooks relation \cite{Crooks98, Crooks99, Crooks00}:
\begin{equation}\label{C}
\frac{P(w)}{\tilde{P}(-w)}=\exp\frac{w-\Delta \Omega^{eq}}{k_B T}.
\end{equation}
A key remark at this stage is that the fluctuating work (\ref{dotw}) naturally stops evolving when the time-dependent perturbation stops. The ending state can therefore be any arbitrary nonequilibrium state and the final value of the grand potential will correspond to the value of the time-dependent perturbation at that final nonequilibrium state. The Crooks relation thus only requires the initial condition to be in an equilibrium state.

We next turn to another relation that derives from (\ref{dft}), irrespective of whether or not $\tilde{\tilde{\bold{P}}}=\bold{P}$. Indeed we have: 
\begin{equation}
\int \;d\Delta_i{s}\; e^\frac{-\Delta_i{s}}{k_B}\;{P(\Delta_i{s})}=\int\; d\Delta_i{s}\;{\tilde{P}(-\Delta_i{s})}=1,
\end{equation}
where we use the fact that $\tilde{P}$ is a probability density, hence normalized to one.
We thus obtain the so-called integral fluctuation theorem, which is valid irrespective of any conditions on initial or final states \cite{Seifert05}:
\begin{eqnarray}\label{ift}
\langle \exp\frac{-\Delta_i{s}}{k_B} \rangle=1.
\end{eqnarray}
If the initial state corresponds to equilibrium (for the same reason as the Crooks relation the final state need not be at equilibrium), we obtain from (\ref{entropywork}) the famous Jarzynski relation \cite{Jarzynski97, Jarzynski97b} (usually written in the absence of particle exchange, with the Helmholtz free energy $F$ replacing the grand potential $\Omega$):
\begin{eqnarray}\label{jarzynski}
\langle \exp\frac{-w}{k_B T}\rangle=\exp\frac{-\Delta \Omega^{eq}}{k_B T}.
\end{eqnarray}
 
As a closing remark, we mention that the integral fluctuation relation and a fortiori the detailed fluctuation relation imply the positivity - on average - of the entropy production:
\begin{eqnarray}\label{ift}
\langle \Delta_i{s}\rangle\geq 0.
\end{eqnarray}
This result can be reproduced in a more direct and revealing way from  the result (\ref{ept}) for the stochastic entropy production. Its average is given by:
\begin{eqnarray}
\langle \Delta_i{s}\rangle=k_B D({\bold{P}(\bold{m})}||{\tilde{\bold{P}}(\tilde{\bold{m}})}),
\end{eqnarray}
where D is the relative entropy introduced earlier:
\begin{eqnarray}
D({\bold{P}(\bold{m})}||{\tilde{\bold{P}}(\tilde{\bold{m}})})=
\int_{\bold{m}}\;d{\bold{m}}\;{\bold{P}(\bold{m})} \ln\frac{\bold{P}(\bold{m})}{\tilde{\bold{P}}(\tilde{\bold{m}})}.
\end{eqnarray}
This quantity is zero, if and only if ${\bold{P}}({\bold{m}})=\tilde{\bold{P}}(\tilde{\bold{m}})$, $\forall {\bold{m}}$. This reveals again the stringent conditions associated to the absence of entropy production, namely the vanishing of any time-asymmetry: every single trajectory and its reverse have to be equally probable. This observation is in accordance with the founding principle of the second law, namely the absence of a perpetuum mobile of the second kind. Indeed, any time-asymmetry in a trajectory could in principle be used to extract work, hence  the system cannot be at equilibrium under such a condition.
 
\section{Discussion}

Stochastic thermodynamics offers a good marriage between statistical physics and thermodynamics. We described the  system via its microscopic energy states, with Markovian dynamics induced by its contact with idealized work and reservoirs. One may wonder whether the obtained results are limited to this particular setting, or are in fact much more general and deep. As mentioned in the introduction, many of the results, and in particular the detailed and integral work and fluctuation theorems have been obtained in other settings lending credance to the believe that we are indeed dealing with a more profound reformulation of thermodynamics. The further development and consolidation of this theory is in fact in full swing, including surprising findings such as integral and detailed fluctuation theorems for quantities other than the entropy production. We have not discussed here the microscopic origin of irreversibility: the Master equation is irreversible from the very start. Nevertheless, the picture presented by stochastic thermodynamics is fully consistent with a microscopic derivation that addresses this issue, see in particular \cite{EspoLindVdBNJP10}.

\acknowledgments

This research is supported by the research network ``Exploring the Physics of Small Devices'' of the European Science Foundation. C. V.d.B. is supported by the ``Fonds voor Wetenschappelijk Onderzoek Vlaanderen'', and M. E. by the ``National Research Fund of Luxembourg'', project FNR/A11/02 and INTER/FWO/13/09.

%
\end{document}